\documentstyle[psfig]{caosp}

\def\Fe2{\hbox{{\rm Fe}~{$\scriptstyle {\rm II}$}}}
\def\Cr2{\hbox{{\rm Cr}~{$\scriptstyle {\rm II}$}}}
\def\kms{\,km\,s$^{-1}$}
\def\degr{\deg}
\begin{document}

%\htitle{$\iota$~Cas...}
%\hauthor{G.A. Wade~{\it et al.}}

\title{$\iota$~Cas: Multi-element Doppler imaging and magnetic field geometry}

\author { R. Kuschnig\inst{1}, G.A. Wade\inst{2}, G.M. Hill\inst{3}, N.
Piskunov\inst{4}} \institute{Institut f\"ur Astronomie, Universit\"at
Wien, T\"urkenschanzstr. 17, A-1180 Vienna, Austria \and Physics \&
Astronomy Department, University of Western Ontario, London, Ontario,
Canada, N6A 3K7 \and
           McDonald Observatory,
           University of Texas at Austin,
           P.O. Box 1337,
           Fort Davis, Texas, USA 79734
\and
Uppsala Astronomical Observatory, Box 515, S-751 20 Uppsala, Sweden}
\maketitle

\begin{abstract}
In order to clarify the role of the magnetic field in generating
abundance inhomogeneities in the atmospheres of Ap stars, we present new
abundance Doppler images and an approximate magnetic field geometry for
the Ap star $\iota$~Cas. 

\end{abstract}


\section{Introduction}

 The inhomogeneous distributions of surface chemical abundance
 observed in the Ap and Bp stars are thought to result from a complex
 interplay between gravitational and radiative diffusion,  mass loss,
 turbulence and circulation processes in their atmospheres (Michaud \&
 Proffitt 1993). By determining the magnetic field geometries and
 surface chemical abundance distributions of a representative sample
 of these objects, important new contraints can be placed upon the
 manner and degree to which these processes interact with the magnetic
 field.

 $\iota$~Cas (HD 15089) is classified as A5p in the {\em Henry Draper
 Catalogue}. Its projected rotational velocity is moderately high
 ($\sim$ 50~\kms), making it an ideal candidate for Doppler imaging.
 Borra \& Landstreet (1980) obtained four Balmer-line magnetometer
 measurements of the longitudinal magnetic field of this star, with
 $1\sigma$ uncertainties around 150~G. All but one are consistent with 
 zero field, indicating that the magnetic field is
 quite weak. The photometric period (which we assume to be the rotational
 period) has been previously reported by a number of authors. 

\section{Observations}
%\vspace{-2mm}
 10 spectra of $\iota$~Cas were obtained at
 Observatoire de Haute-Provence using the AUR\'ELIE spectrograph, during
 1994 and 1995.  The spectral resolution of these data is about 
 $2\times 10^4$, and the SNR is about 200:1.

\vspace{-2mm}
\begin{table}
\small
\begin{center}
\begin{tabular}{||l|c||}
\hline\hline
Ephemeris & $2437247.704 + 1.7405\cdot {\rm E}$ \\
$v\sin i$ & 48~\kms\\
Inclination &  $50\degr$\\
$T_{\rm eff}$ & 8500~K\\
$\log g$  & 4.0\\
\hline\hline
\end{tabular}
\end{center}
\caption{Input data for the abundance Doppler images}
\end{table}
\vspace{-3mm}
%\subsection{Longitudinal field}

  15 measurements of the longitudinal field variation of $\iota$~Cas were
  obtained using the photoelectric polarimeter at The University of
  Western Ontario Elginfield Observatory. A detailed description of
  the instrument and the observing technique are given by Landstreet
  (1980). 

\vspace{-3mm}
\section{Magnetic field geometry}
\vspace{-2mm}
  The new magnetic field measurements, phased according to the ephemeris
  cited in Table 1, are shown in Fig. 1. The longitudinal field clearly 
  undergoes a sinusoidal variation, indicative that the geometry of the
  photospheric magnetic field is predominantly dipolar. Assuming such a
  field configuration, and the rotational
\begin{figure}[hbt]
\centerline{\psfig{figure=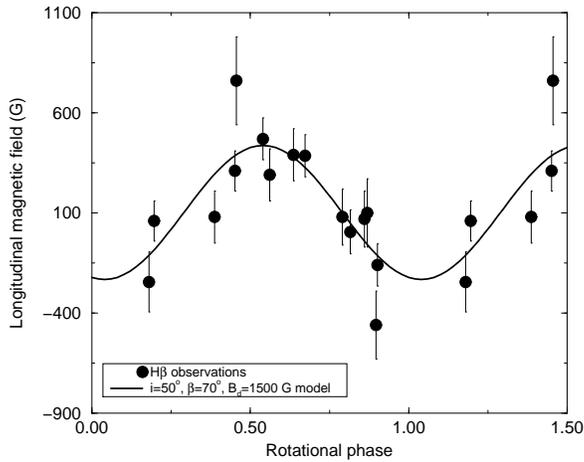,height= 6.1cm}}
\caption{Longitudinal magnetic field variation of $i$~Cas.}
\end{figure}
\vspace{-2mm}
  axis inclination assumed for
  the Doppler images (cited in Table 1), we compute the obliquity
  of the magnetic axis to the rotational axis $\beta=70\degr$ and 
  the polar field $B_{\rm d}=1500$~G using the expressions of Preston (1967).
  According to this model, the magnetic poles are at rotational coordinates
  $(l,b)_+=(195\degr,+20\degr)$~and~$(l,b)_-=(15\degr,-20\degr)$.

\section{Doppler Images}
%\vspace{-1mm}
Abundance Doppler images of $\iota$~Cas were obtained using the
INVERS8 Doppler imaging code (Piskunov \& Rice 1993). In Fig. 2 we
show Cr abundance distribution maps (in rectangular and spherical
projection) and line profile fits computed for the line \Cr2
~$\lambda 4558.65$. Maps of the Fe, Mg and Ti distributions show similar
structure. 

The mean Cr abundance of $\iota$~Cas is enhanced by about 1 dex above
solar. A clear ring of enhanced abundance is apparent in the northern
hemisphere. The coordinates of the northern (positive) magnetic pole as
derived above place it within the enhanced ring. The southern (negative)
pole does not appear to be associated with any obvious abundance feature.
This scenario is similar to that observed for HD~153882, in which Fe
appears to be distributed in an enhanced (broken) ring, with the northern
magnetic pole located within the ring (Ryabchikova et al. 1995). 
\begin{figure}
\centerline{\psfig{figure=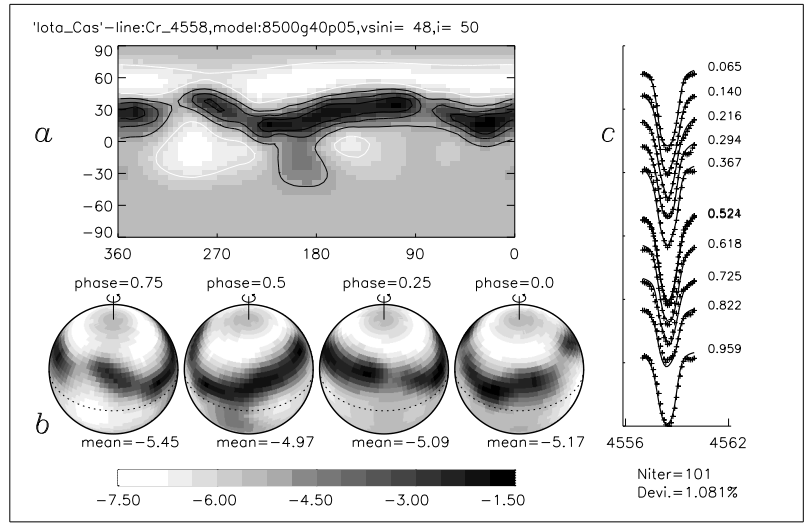,height=7cm}}
\caption{Cr abundance distribution of $i$~Cas.}
\end{figure}

\end{document}